# LG-NuSegHop: A Local-to-Global Self-Supervised Pipeline For Nuclei Instance Segmentation

Vasileios Magoulianitis[1], Catherine A. Alexander[1], Jiaxin Yang[1] and C.-C. Jay Kuo[1]

[1]*Minh Hsieh Department of Electrical and Computer Engineering, University of Southern California (USC), Los Angeles, CA, USA*

ABSTRACT

Nuclei segmentation is the cornerstone task in histology image reading, shedding light on the underlying molecular patterns and leading to disease or cancer diagnosis. Yet, it is a laborious task that requires expertise from trained physicians. The large nuclei variability across different organ tissues and acquisition processes challenges the automation of this task. On the other hand, data annotations are expensive to obtain, and thus, Deep Learning (DL) models are challenged to generalize to unseen organs or different domains. This work proposes Local-to-Global NuSegHop (LG-NuSegHop), a self-supervised pipeline developed on prior knowledge of the problem and molecular biology. There are three distinct modules: (1) a set of local processing operations to generate a pseudolabel, (2) NuSegHop a novel data-driven feature extraction model and (3) a set of global operations to post-process the predictions of NuSegHop. Notably, even though the proposed pipeline uses no manually annotated training data or domain adaptation, it maintains a good generalization performance on other datasets. Experiments in three publicly available datasets show that our method outperforms other self-supervised and weakly supervised methods while having a competitive standing among fully supervised methods. Remarkably, every module within LG-NuSegHop is transparent and explainable to physicians.

---

## 1 Introduction

Cancer diagnosis from biopsy tissue specimens has been the standard way to tumor detection and grading. Cancerous and healthy cells have distinct molecular profiles which can provide important visual cues to pathologists. Nuclei segmentation is a fundamental task within this diagnosis pipeline, since the nuclei cell topology, size and shape play a crucial role to cancer grade reading. Hematoxylin and Eosin (H&E)-stained images give rise to this molecular profile by highlighting the nuclei cells and it has been the cornerstone process for histolopathological slides preparation [40].

Undoubtedly, histopathological image reading is a painstaking task. It relies on very subtle visual cues, requiring also highly expertise. On top of this, digitized slides are captured usually under a high magnification level, typically ranging from 20x-40x. That results in very high resolution images which pathologists need to examine thoroughly to recognize potentially cancerous regions. Given that multiple cores usually sampled out of each patient, one can realize that analyzing histology slides is a fairly time consuming and laborious task [10]. Computer-aided diagnosis (CAD) tools are meant to automate certain physicians' tasks, offering also a more objective decision making process. Automated nuclei segmentation can expedite the slide reading process by highlighting the molecular patterns and enhance pathologist's reading. Moreover, it can be used as the intermediate step toward whole-slide classification (WSI) for models aiming to learn the pattern of clusters that nuclei form and map that to a grade group of cancer [34, 4].

Nuclei segmentation poses several challenges to models and algorithms. At first, the H&E staining process [35] involves many steps carried out manually from humans and thus it is far from stable and noise-free. Staining artifacts can also increase the intra-class distance, while during the image acquisition process, the type of scanner and its parameterization can also affect the nuclei appearance [54]. Another challenge that modern Deep Learning (DL) models are faced with is the lack of large annotated datasets, since it is a labor intensive task that only expert pathologists can perform. Therefore, data annotation is expensive and also subject to high inter-observer variability [6], which regards this problem as learning from noisy labels.

There is a plethora of works in existing literature which approach the problem from different angles. Prior to the DL-based methods, most of the works focused on unsupervised methodologies. For instance, different variants of thresholding operations [59, 41], active contours [1] and level sets [7], watershed algorithm [17, 32], Graph cuts [16] and K-means clustering [58]. Those approaches were mostly relied on biological priors of the problem, particularly about the nuclei appearance, shape and size.

It has been almost a decade since the advent of DL in the medical imaging field. For segmentation tasks, fully convolutional pipelines, such as U-Net [47] are popular choices among the researchers for semantic segmentation. Fully supervised methods use U-Net as backbone architecture [8, 14], also coupled with attention mechanisms tailored to focusing the learning on the error-prone regions (i.e. nuclei boundaries) [70, 29]. Since fully supervised methods are heavily challenged from the lack of large annotated datasets, weakly supervised methods[67, 27] attempt to learn either using less labels or point-wise annotations [26, 44]. Furthermore, unsupervised learning methods use self-supervision and specifically employ domain adaptation [13] and predictive learning [49] to transfer the nuclei appearance from other domains. Nevertheless, they fail to achieve a competitive performance.

Despite their success in other computer vision problems, DL models are challenged in the medical imaging tasks, mainly due to the lack of large datasets. More importantly, DL models are often criticized as "black-boxes" from physicians [42], since inherently their feature learning process is intricate. Moreover, to achieve a good performance, backbone models require pretraining on the ImageNet. As such, it is unclear how the representations can be adapted from a natural imaging domain to the biological one. Furthermore, those models fail to explicitly incorporate human's prior knowledge which is important for a transparent decision making from the tools.

All the mentioned reasons motivate this work to attempt a fully unsupervised pipeline and also decouple from the DL paradigm. Instead, a novel data-driven feature extraction model for histology images is introduced, namely NuSegHop. It is a linear, feedfoward and multi-scale model to learn the local texture from the histology images. Our approach is based on the Green Learning (GL) [20] paradigm which offers a framework for feature learning at a significantly lower complexity, where the features can be seamlessly interpreted [21]. The proposed pipeline consists of three major modules, starting with a set of local processing operations using priors of the task to generate a pseudolabel. Then, NuSegHop is used in a self-supervised manner to predict a heatmap for nuclei presence. Finally, a set of global processing operations takes place as a post-processing to decrease the false negative and positive rates, also in a self-supervised manner. Overall, the full pipeline incorporates self-supervised learning and priors insights, ranging from local areas (i.e. patches), up to global image decisions. Therefore, the overall proposed pipeline is named Local-to-Global NuSegHop (LG-NuSegHop).

The main contributions of this work are:

- NuSegHop as a data-driven feature extraction model to learn the texture in histology images in an unsupervised way.

- LG-NuSegHop self-supervised pipeline which combines local and global image processing techniques, along with the NuSegHop model for nuclei segmentation.

- Competitive performance in three diverse datasets among other DL-based supervised and weakly supervised models. Also, extensive quantitative and qualitative comparisons and discussions are offered to help realizing in which cases supervision makes a difference in this task.

- Cross-domain experiments showing high generalization performance across multi-organ datasets, with different staining methods.

## 2 Related Work

In this section we provide an overview of methods about nuclei segmentation across different categories, beginning with the traditional pipelines that our work has elements from, and further including the DL state-of-the-art methods. For a more comprehensive and detailed overview of nuclei segmentation, a recent survey [35] conducted that provides detailed explanations and comparisons.

### 2.1 Traditional Methods

Earlier works relied mostly on priors from the nuclei appearance and certain assumptions to solve the problem. Thresholding was fundamental in early segmentation works, where different methods propose mechanisms for calculating the appropriate threshold to binarize the input image. A popular algorithm in many works is Otsu's thresholding [41, 2], trying to minimize the intra-class variance or maximizing the inter-class one and automatically discover the best threshold. Win *et.al.* [59] apply a median filter on each color component and then perform Otsu's thresholding on the grayscaled image, followed by morphological operations to refine the output. A locally adaptive thresholding mechanism on linear color projections has been also proposed in [38].

Watershed algorithm [46] is another popular approach that uses topological information to segment an image into regions called catchment basins. This algorithm requires initial markers which are the seeds of the catchment basis. It is a popular choice in many works a post-processing step to find the nuclei boundaries [32, 60].

### 2.2 Learning-based Methods

#### 2.2.1 Full Supervision

The initial DL-based works [62] fully relied on pathologists' annotations to learn the nuclei color and texture variations. Region-proposal works employ Mask-RCNN to detect the nuclei [24, 48]. One of the most popular architectures for medical image segmentation used as a backbone in several works is the U-Net [47]. Kumar *et al.* [19] proposed a 3-way CNN model trained to supervise the nuclei boundaries.

Instead of a binary map, it produces a ternary one which helps distinguishing better the nuclei from background. Essentially, this study shows that attention on the nuclei boundary improves the accuracy of the detection and segmentation.

CIA-Net [70] leverages the mutual dependencies between nuclei and their boundaries across different scales, and combines two distinct decoders, one for nuclei and one for their contours. It also proposes the Truncated Loss for diminishing the influence of outlier regions and mitigate the noisy labels effect. Moreover, to enhance the multiscale learning capabilities, it introduces lateral connections between the encoder and decoder in each layer. In this way, the texture information learned in the early layers can be combined with the semantic features from the deeper ones. Graham *et al.* [8] introduce Hover-Net, a multi-branch network that is trained on segmentation, classification and pixel distance from the nuclear mass targets. In the loss function, the mean squared error (MSE) is calculated between the ground truth and the distance map, as well as the MSE of the distance gradients. By including the gradients into the loss function, it was found that it helped to delineate the nuclei boundaries more accurately.

As emphasized, full label collection in this task is expensive and not in abundance. To this end, point-wise labels [44, 66] can be used to learn the appearance of nuclei from partial point annotations. Furthermore, it is possible to combine point annotations and a limited number of full nuclei masks to enhance the learning process and improve the results [45].

### 2.2.2  Self-Supervision

Several methods have used self-supervision, to learn from a different task domain and transfer the knowledge into nuclei segmentation. Domain adaptation is a popular self-supervised choice since it exploits the large volumes of labeled data from other domains, and then apply it to the target domain. Domain Adaptive Region-based CNN (DARCNN) is proposed in [13] which learns definition of objects from a generic natural object detection dataset [25] and adapts it on the biomedical datasets. This is possible through a domain separation module that learns domain specific and domain invariant features. Liu *et al.* [28] propose the Cycle Consistent Panoptic Domain Adaptive Mask RCNN (CyC-PDAM) that learns from fluorescence microscopy images and synthesizes H&E stained images using the CycleGAN [71]. Contrastive learning is another way for applying self-supervision. Xie *et al.* [61] propose an instance aware self-supervised method which involves scale-wise triplet learning and count ranking to implicitly learn nuclei from the different magnification levels.

Predictive learning is another alternative to learn representations implicitly from the data. Sahasrabudhe *et al.* [49] have proposed a method on the assumption that the image magnification level can be determined by the texture and size of nuclei. In turn, this can be used as a self-supervised signal to detect nuclei locations and seed the watershed algorithm. Zheng *et al.* [68] have proposed a method that generates pseudo labels obtained from an unsupervised module using k-means clustering on the Hue-Saturation-Intensity (HSI) colorspace. Then, an SVM classifier is trained on a feature vector with color and texture, as well as topological attributes. Our work is conceptually similar to that work, since it creates a pseudolabel from local thresholding operations and then uses self-supervision at a global level.

### 2.3  Green Learning

Green Learning (GL), has been recently introduced in[20], aiming to provide a more transparent feed-forward feature extraction process, at a small complexity and model size [21, 64]. The proposed feature extraction model create a multi-scale feature extraction and creates a rich spatial-spectral representation of the input image [5]. Instead of convolutional filters trained with backpropagation, principal component analysis (PCA) is used to learn the local subspace across different layers, where each feature has larger receptive field along deeper stages. Following GL's terminology, each layer is called "Hop", in which features are learned in an unsupervised and data-driven way.

Within the medical imaging field, Liu *et. al.* [31, 30] have proposed the first works on segmentation and classification tasks. Also, GL has recently achieved competitive results in prostate cancer detection from Magnetic Resonance Images [37]. From the same framework, a U-Net inspired model, namely GUSL, has been introduced for medical image analysis segmentation problems, with applications to prostate gland segmentation. It is a fully supervised method, meant to offer multi-scale feature extraction and semantic segmentation at different scales, introducing a novel coarse-to-fine scale regression model. Our core proposed module, NuSegHop, uses the channel-wise Saab transform from GL for pixel-wise feature extraction and classification in a self-supervised manner. To the best of our knowledge, this is the first GL-based work in digital histology.

## 3 Materials And Methods

Although the existing DL-based methods usually comprise one model that is trained in an end-to-end manner using backpropagation, LG-NuSegHop has distinct modules, and each of those has a discrete task within the pipeline. This approach decouples from the DL paradigm, and one of its key benefits is the transparency of every step, since every module has a specific task and role within the overall pipeline. The proposed pipeline comprises three distinct modules that operate successively. In Section 3.1, we describe the image preprocessing steps, meant to enhance the input image towards the subsequent operations. In Section 3.2, the local pixel-wise operations are described to predict a pseudolabel for NuSegHop, where its architecture and process is detailed in Section 3.3. The global processing modules are presented in 3.4. An overview of the entire proposed LG-NuSegHop pipeline is illustrated in Fig. 1.

### *3.1 Preprocessing*

The preprocessing modules aim at preparing the input image tile for the subsequent local processing module which is mainly based on thresholding. Thus, the key goals are: (1) highlight the nuclei over the background and (2) convert the color image into grayscale.

Prior to thresholding, the goal is to make nuclei more distinct over the background tissue. From the theory of the H&E staining process, hematoxylin principally colors the nuclei cells to a darker color (e.g. blue or dark-purple), while eosin mainly stains the cytoplasm and other structures in the background area. There are several methods in literature for carrying out this color conversion. We choose the work of Salvi *et al.* [50]. It is an Singular Value Decomposition (SVD)-geodesic based method for stain separation, after converting the input image from RGB to the optical density space, where SVD can be more effective. Another benefit from stain separation, it helps mitigating the large stain variability across different images which is one of the challenges in this task.

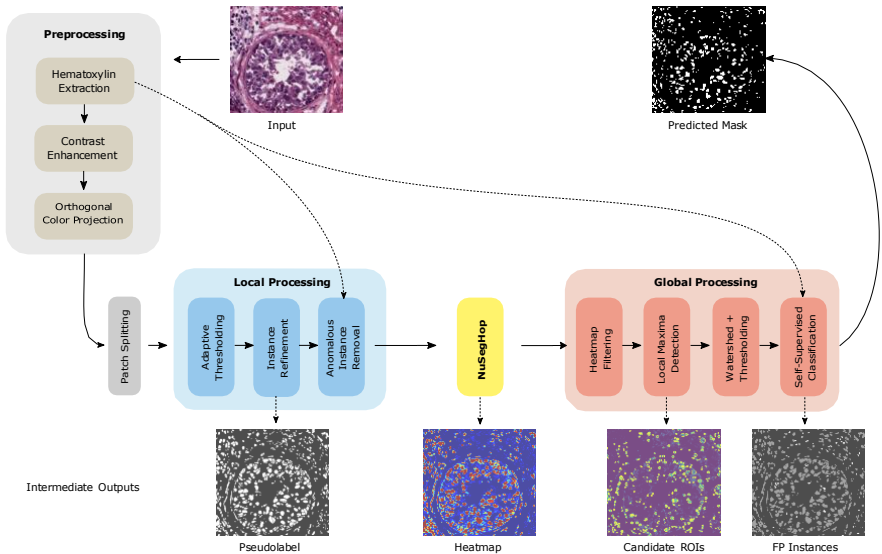

Figure 1: An overview of LG-NuSegHop pipeline. Pre-processing applies image enhancement to prepare the image for the local processing modules. NuSegHop receives as input the pseudolabel and predicts a heatmap. In the last step, global processing modules increase the nuclei detection rate using information across the entire input image.

After separating the stain colors using the orthogonal spaces from SVD, we project all pixels on the Hematoxylin's subspace -eigenvector from SVD corresponding to nuclei- to create image $H$. To further enhance this separation and make also nuclei boundary more distinct – especially for images that suffer from blurry artifacts due to the staining or acquisition process– we apply histogram equalization.

The last preprocessing step to prepare the input tile for the thresholding operation is convert it to grayscale. Although other colorspaces (e.g. LAB) could be an option, in our earlier work [36] we have shown a more optimal way to convert the image into grayscale. Transformation across different colorspaces use certain formula to map pixel values from one domain to the other. PQR, named after the three channels of principal component analysis, is a data-driven way for color conversion, adapted to the input content. It is based on SVD to calculate the color subspace direction that maximizes the data variance. One advantage is the better energy compaction in one channel, comparing to a fixed colorspace conversion. Moreover, finding the color conversion that maximizes the variance is particularly important for the subsequent thresholding operation, since we assume that along the direction of that subspace, the separation between nuclei and background is maximized. Therefore, after SVD, we linearly project patches $H_p$ from the H image on its first principal component P (p-value) to convert from color to grayscale. Since, different areas of the tile may have different statistics, we perform PQR independently after the image is split in local patches which are subject to thresholding. An illustrative example of PQR is shown in Fig. 2. The color conversion formulae are as follows:

$$H_p = U \cdot S \cdot V^\top \qquad (1)$$

$$P \triangleq V_{1,1:}, \quad Q \triangleq V_{2,1:}, \quad R \triangleq V_{3,1:} \qquad (2)$$

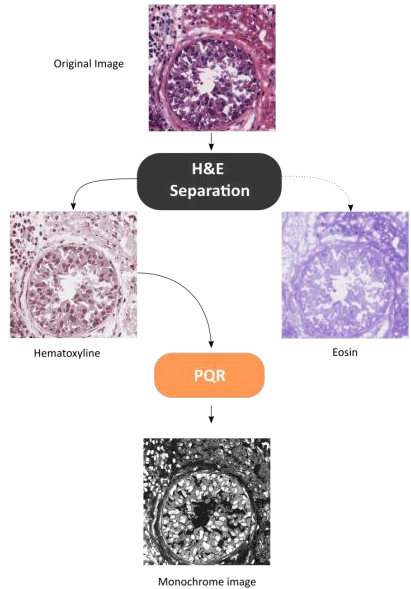

Figure 2: An illustration of the main preprocessing steps, involving stain separation and the PQR method to convert color into grayscale.

### 3.2 Local Processing

The main purpose of this module is to create a pseudolabel for training NuSegHop. This module uses simple, yet effective and intuitive image processing techniques at a local level. It employs prior knowledge of the problem and self-supervision locally, to filter out error predictions from the unsupervised local processing. To this end, certain assumptions are made, to overcome the lack of supervision:

1. The bi-modal distribution according to which at a local area histogram there are two peaks, where the lower intensity corresponds to nucleus and the brighter to background. Preprocessing is meant to accentuate this assumption

2. Local similarity where adjacent nuclei tend to have less color or texture variations

3. Larger low intensity components are less likely to be false positives than the smaller instances

#### 3.2.1 Adaptive Thresholding

The thresholding method we propose is adaptive in two ways: (1) scale-wise and (2) intensity-wise. The input image is split into patches of size $P$ $50 \times 50$ and the process starts out with estimating the local distribution. If the bi-modal criteria are not met, the process tries also patches of $25 \times 25$ and $100 \times 100$. This is to adapt on different nuclei sizes or magnification levels. On the other hand, the threshold at each local patch is automatically adjusted based on the local area statistics. One choice for threshold calculation is to simply pick the intermediate value between the two peaks. Yet, we opt for a more adaptive way to calculate the optimal threshold [38], thus reducing the under or over segmentation effects and eventually create a less noisy pseudolabel for self-supervision. Given the histogram and the two main peaks $T_1$ and $T_2$ under the bi-modal assumption, we define as $L_{12}$ the line crosses through $T_1$, $T_2$, the intermediate point $T_o$ where the threshold correction is about $T_o = \frac{T_1+T_2}{2}$. Also, $T_c$ is defined as the intercept point of the intensity value axis of the histogram and the perpendicular line of $L_{12}$, passing from $T_o$ (see Fig. 3). By using this heuristic and simple method to correct the intermediate p-value from (P)QR pre-processing, the local thresholding is more resilient to false positive and negative pixels. More details can be found in [38]. A $\lambda$ hyperparameter is used to control the amount of correction about $T_o$. The adjusted threshold $\hat{T}$ formula is calculated using Eq. 3.

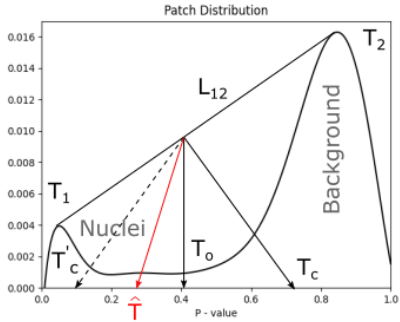

Figure 3: Demonstration of the P-value distribution in a local patch where the bimodal assumption holds. The auxiliary lines to calculate the adapted threshold $\hat{T}$ are also depicted.

$$\hat{T} = T_o + \lambda(T_o - T_c) \tag{3}$$

### 3.2.2 Morphological Instance Refinement

Although adaptive thresholding works well in areas with relatively low variation, there are patches where the color variance is higher, thereby causing over or under segmentation effects. Morphological processing has been widely used in literature for processing binarized images. To refine the thresholding operation, we apply a set of morphological operations, such as hole filling, small instance removal and nuclei splitting.

Priors about the nuclei size and shape are incorporated to apply simple morphological processing and filter out noisy instances. For nuclei splitting, the convex hull algorithm is employed to detect highly deep curvatures that are not indicative of nuclei shape. This step is significant since subsequent operations operate on a per-instance base.

### 3.2.3 Locally Anomalous Instance Removal (LAIR)

To further filter out larger instances that are possible to be false positives, we carry out a simple local comparison among the detected instances. For this operation we need a larger patch, in order to include more nuclei instances and make the comparison effective. Hence. for this submodule a 200 × 200 patch is used. In each large patch, the first criterion for query instances $q$ is the size. That is, if an instance has a small to medium size, it will be compared against the rest larger instances $r$ that provide reference, according to assumption 3 (see Section 3.2). We create a reference representation by forming the ensemble from the non-query instances. This can be viewed as a first step of introducing self-supervision locally in our pipeline, coupled with priors from the task. Intuitively, abnormally looking instances at a certain feature space can be regarded as anomalies and in turn eliminated. We define, $Q = \{q_1, q_2, \ldots, q_N\}$ the instances being tested and $R = \{r_1, r_2, \ldots, r_M\}$ the reference instances.

Regarding the feature representation, we use the HSI colorspace from the $H$ image, along with the channel-wise contrast value. For similarity comparison, a Gaussian kernel is used to measure the distance between each query instance $x_j$ and the ensemble reference $x_R$ (see Eq. 4), and determine the anomaly instances that are subject to removal obtaining a similarity score $S$ (see Eq. 5). Instances $q$ that have a lower similarity with their local reference class $R$ from a predefined threshold $T_s$ are removed from the foreground.

$$\mathbf{x}_R = \frac{1}{|R|} \sum_{i \in R} \mathbf{x}_i, \tag{4}$$

$$S(\mathbf{x}_R, \mathbf{x}_j) = e^{-\gamma ||\mathbf{x}_j - \mathbf{x}_R||^2_2}, \quad \forall j \in Q, \tag{5}$$

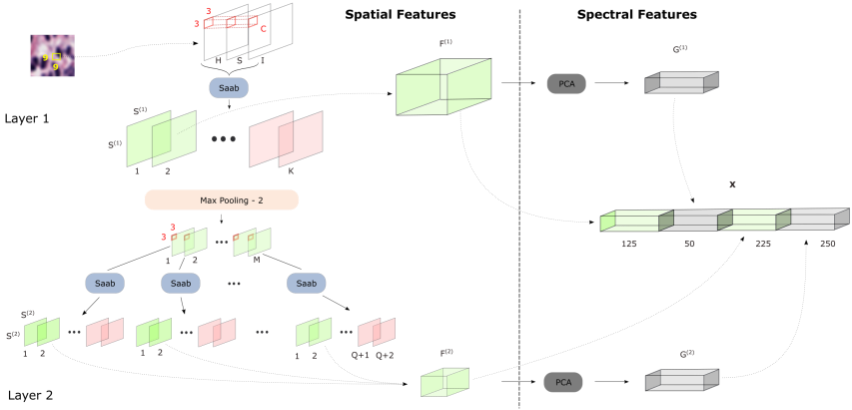

Figure 4: Graphical overview of the proposed NuSegHop for feature extraction. It consists of two layers that operate in two different scales. From both layers two types of features are extracted: (1) spatial and (2) spectral. With red we depict the extracted feature maps that have low energy and will be discarded. All the spatial feature maps (green color) are concatenated to extract the spectral ones (gray color).

### 3.3 NuSegHop

After the local processing module operations, we have obtained an initial segmentation output using no labels or training data. This output can be used as a pseudolabel to a classifier. Aiming at obtaining a probability heatmap for nuclei segmentation, we propose a novel and unsupervised feature extraction method, named NuSegHop, to learn the local texture of nuclei for pixel-wise classification. Other methods in the past [68] have used hand-crafted features or pure color-based methods which lack robustness. A data-driven and multi-scale feature extraction is proposed in this paper for pixel-wise nuclei segmentation using the GL paradigm [21]. This method has as a key advantage the low complexity and small model size, which is essential for fast inference. Moreover, the GL-based feature extraction module is linear and thereby more transparent and interpretable [64]. This is another major distinction from the DL models that use backprogagation for learning feature representations. NuSegHop learns features in a feed-forward manner and does not require large datasets in order to obtain robust feature representations. Therefore, GL-based methods have certain advantages in specific tasks and this is why we opt for such a feature extraction method. A detailed architecture of NuSegHop is illustrated in Fig. 4.

For NuSegHop, the $H$ image is converted into the HSI colorspace before feature extraction. Originally, a window area $A$ of size $S^{(1)} \times S^{(1)}$ is considered to characterize the area about a pixel. For this problem, we choose $S^{(1)} = 9$, since too small windows may not be able to learn the local texture, while larger ones may induce more noise into feature learning. NuSegHop learns the texture within the window area in two scales $S_1$ and $S_2$, to give multi-scale properties in the feature space. The core operation in NuSegHop for texture learning is the Saab transform [21], which is based on the Principal Component Analysis (PCA), applied in two ways: (1) spatial and (2) spectral. The full feature extraction diagram of NuSegHop is shown in Figure 4.

### 3.3.1 Feature learning - Spatial Saab

To learn the texture across different local areas within a window, a neighborhood construction with filters of spatial size of $f^{(1)} \times f^{(1)}$ is applied with stride 1 and equal padding. Since the input image has three color channels, each local neighborhood defines a cuboid $C^{(1)}$ with size $K^{(1)} = f^{(1)} \times f^{(1)} \times 3$ which contains the local HSI color information. As a consequence, $K^{(1)}$ is the maximum number of extracted subspaces from the Saab transform in layer 1. $L^{(1)} = S^{(1)} \times S^{(1)}$ such cuboids can be extracted from a window at layer-1 using padding on $A$. By sampling across windows centered on the pixels of the original image, one can create a tensor $T^{(1)}$ for training with size $N \times L^{(1)} \times K^{(1)}$, where $N$ the number of sampled pixels from the input image to training NuSegHop. In turn, $T^{(1)}$ can be used for training layer-1 and calculate the subspaces (Eq. 6) used for feature extraction. After SVD matrix decomposition, the rows of $V$ are the eigenvectors correspond to the orthogonal subspaces of the signal (see Eq. 7).

$$T^{(1)} = U \cdot S \cdot V^\top \qquad (6)$$

$$W_{(1)} = V_{1:M, 1:} \qquad (7)$$

In training, Saab transform is applied on $C^{(1)}$ to extract $K^{(1)}$ orthogonal subspaces (i.e. principal components). Moreover, because many principal components may carry no significant energy –as it is dictated from their corresponding eigenvalues–, they can be discarded to remove unnecessary complexity and noise. That is, the first $M$ principal components are retained, with $M < K^{(1)}$.

We define the weight matrix $W^{(1)} \in R^{M \times K^{(1)}}$ which contains all the information to decomposing the cuboids into their spectral representations by projecting onto the extracted principal components. After training, $W$ incudes the weights for feature extraction. For instance, to perform spatial feature extraction in layer 1, one needs to multiply $L^{(1)}$ cuboids extracted within $A$ window, and project them along the $M$ principal components. To formulate this operation, we construct matrix $C$ with size $L^{(1)} \times K^{(1)}$, where its rows contain the cuboids. By multiplying them, we calculate matrix $F$ of size $L^{(1)} \times M$ which includes all the spatial features of layer-1 (see Eq. 8). Spectral maps $F^{(1)}$ in layer 1 are obtained from the matrix $F$, by reshaping it back to size $S^{(1)} \times S^{(1)} \times M$. We view each principal component as a different spectral local representation of $A$.

$$F = C \cdot W^\top \quad (8)$$

We also concatenate as additional feature (one more column in $F$) the mean of each local cuboid in both layers (Eq. 9), since apart from the texture, the local color is also important to differentiate nuclei from background. If we are to draw a parallel with circuit theory, the DC component is the mean color and AC are the textures derived from the Saab transform. Thus, $F$ is now of size $L^{(1)} \times (M + 1)$.

$$F = F_{DC} \oplus F_{AC} \quad (9)$$

In layer-2, feature map $F^{(1)}$ is fed as input after a max-pooling layer. The spatial feature extraction process in layer-2 is similar to layer-1, with one difference, the Saab transform is applied independently on each of the $M+1$ feature maps of layer-1 [5]. Therefore, after neighborhood construction each cuboid $C^{(2)}$ has a shape of $K^{(2)} = f^{(2)} \times f^{(2)} \times 1$. Also, the spatial size of $L^{(2)} = S^{(2)} \times S^{(2)} = \lceil (S^{(1)}/2) \rceil \times \lceil (S^{(1)}/2) \rceil$ after max-pooling. The filter sizes and output shapes of each layer are show in in Table 1.

Table 1: Architecture of the proposed NuSegHop.

|  | F Resolution | f Size | Stride |
| --- | --- | --- | --- |
| Layer 1 | $(9 \times 9) \times 3$ | $(3 \times 3) \times 1$ | $(1 \times 1) \times 1$ |
| Max-pool 1 | $(5 \times 5) \times 1$ | $(2 \times 2) \times 1$ | $(2 \times 2) \times 1$ |
| Layer 2 | $(5 \times 5) \times 1$ | $(3 \times 3) \times 1$ | $(1 \times 1) \times 1$ |

Independent Saab transforms as many as $M + 1$ are applied on tensors with size of $N \times L^{(2)} \times K^{(2)}$. After channel-wise Saab transform in layer-2, each spectral map originally has a size of $S^{(2)} \times S^{(2)} \times (M + 1) \times K^{(2)}$, after concatenating all the feature maps from the channel-wise Saab transforms. Energy-based spectral truncation is also applied in layer-2. Supposing that $Q$ principal components are kept from each channel-wise Saab transform ($Q < C^{(2)}$), then the final layer-2 spectral maps will have a shape of $S^{(2)} \times S^{(2)} \times (M + 1) \times (Q + 1)$, adding also the DC channels in the same way as in layer 1. The energy threshold for both layers is set at $T_e = 1e - 03$.

### 3.3.2 Feature learning - Spectral Saab

Spatial-wise Saab provides a spectral analysis of $A$ across all its spatial regions at the scales of $S^{(1)}$ and $S^{(2)}$ $A$. Therefore, each feature has a spatial correspondence. Yet, it is required to extract features that have a global reference to $A$ as well. Those features are unassociated from the spatial domain and meant to capture different patterns within $A$ area, such as boundary transitions from nuclei to background. To this end, on each spectral map $F^{(1)}$ and $F^{(2)}$ from layers 1 and 2, we apply a PCA using the spectral maps' spatial components as features. By doing so, the transformed signal will have no spatial correspondence anymore. This is performed independently for every $M + 1$ and $(M + 1) \times (Q + 1)$ spectral maps for layers 1 and 2, respectively. The spectral features $G_s^{(l)}$ from each layer is simply the union of all PCA transformed spatial features $F_s^{(1)}$. (see Eq. 10 and 11.) The same $T_e$ is used to filter out the principal components and reduce the dimensionality.

$$G_s^{(1)} = \cup_{s=1}^{M+1} \{PCA(F^{(1)})_{T_e}\} \qquad (10)$$

$$G^{(2)}_s = \bigcup_{s=1}^{M+Q+2} \{PCA(F^{(2)}_s)_{T_e}\} \quad (11)$$

The last step in NuSegHop is to concatenate all the spatial and spectral features from both layers to form the final feature *X* that characterizes *A*. After concatenation the top 100 discriminant features are selected [65]. This provides a rich spatial-spectral feature representation about the color and texture of the local neighborhood under *A*. Besides, in the Saab feature space, the spectral dimensions are uncorrelated because the principal components are orthogonal by definition.

NuSegHop enables for fast pixel-wise predictions, requiring no supervision for its feature extraction part. Having extracted the features on each *W*, one can train a classifier using the pseudolabels from the local processing module. We train an Xtreme Gradient Boosting (XGB) classifier and use its probability predictions to generate a heatmap $\hat{P}$. Each pixel contains the probability of belonging to the foreground.

In the LG-NuSegHop pipeline each module has a clear scope for improving certain aspects of the nuclei segmentation task. Therefore, each module's purpose and output are transparent. The linearity property in NuSegHop features enables the visualization of the local patterns that weigh in classifying a pixel. Concretely, during inference, one can trace back the more "informative" NuSegHop feature dimensions of the classifier. In turn, those features can be mapped back to the input layer –using the inverse Saab transform–, to elucidate the texture and color patterns they correspond to. In other words, NuSegHop enables the pathologist to review the visual elements that classify a certain region as nuclei or background. This seamless decision interpretation provides great advantages in clinical applications since pathologists can understand the decision-making process of the tool, thereby making it more trustworthy for clinical deployment.

### 3.4   Global Processing

This module aims at integrating the locally made decisions, based on color and texture, and perform a global post-processing. Most operations from the local processing group are pixel-wise and carried out in local patches of the original image to reduce variability, whereas global processing has the entire information about the image.

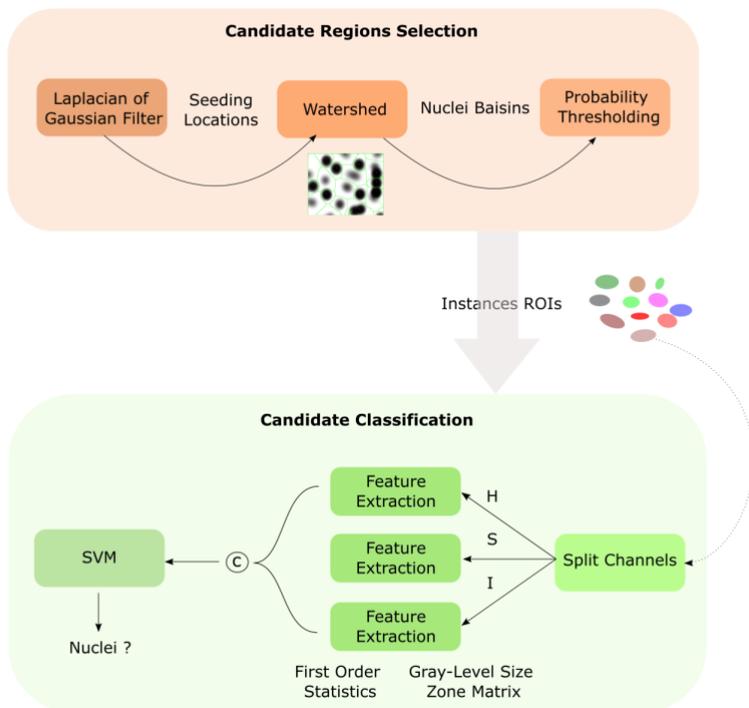

Figure 5: An overview of the global processing pipeline. The Laplacian of Gaussian filter detects local minima from the NuSegHop heatmap to decrease the false negative rate. Watershed and probability thresholding binarize the image and delineate nuclei boundary from the heatmap. Candidate instances are classified in a self-supervised manner to detect any false positives.

As mentioned, locally-based decisions may miss faintly stained nuclei or misclassify background areas as nuclei. The goal of moving from local to global decisions is to decrease the false negative (FN) rate by including more instances as candidates, based on the probability areas from the local decisions. Since, it is inevitable for this process to give rise to false positives (FP), self-supervision is also employed at the end of the global processing modules to help discern potentially FP instances. A diagram of the global processing pipeline is shown in Fig. 5.

### 3.4.1 Heatmap Filtering

Out of our observations on the obtained probability heatmap $P$, most of the nuclei are predicted with high confidence from NuSegHop module. This refers to solid stained instances that can be easily recognized from their color and texture. As we want to decrease the complexity of the global processing unit, large instances are retained and we consider only the less confident ones for the subsequent module (i.e. local maxima detection). This helps both the complexity and efficiency. To do so, the heatmap and the predicted mask from NuSegHop are considered to calculate the per instance confidence, by taking the average from all pixels belong to the instance. Highly confident instances with average probability more than 0.95 are removed from the local maxima detection submodule. After this submodule we obtain the filtered heatmap $P'$.

### 3.4.2 Local Maxima Detection

This submodule aims mainly at increasing the recall ratio of nuclei, on the remaining areas after instance filtering, where the NuSegHop unit is not confident. Texture and color variations or faintly stained nuclei from the local processing module result in scattered high probability areas that during binarization become isolated small instances. The goal is to detect those areas and create candidate ROIs as foreground. Given the filtered heatmap $P'$ this task boils down to a local maxima detection (LMD). We apply a Laplacian of Gaussian filter (LoG) to detect "blob-like" regions which corresponds to potential nuclei instances. The Gaussian filter is meant to smooth and unify the pixel-wise heatmap estimation, thus to mitigating the color and texture variance.

### 3.4.3 Watershed Post-processing

For the highly confident instances, their boundary is typically distinct and can be determined from the local processing modules and in turn from the heatmap. However, for the low confidence instances that are hard to be accurately classified based on the heatmap. We can detect the position of the nuclei but it is hard to estimate accurately their boundaries, since those areas are outliers when training the classifier.

As long as the rough locations of candidates are obtained from the previous module, we their centroids to seed the watershed algorithm, and find the adjacent nuclei boundary lines. This helps areas with multiple less confident nuclei located therein, where their boundaries estimation is more challenging. As a last step, we binarize the filtered heatmap, appending back also the confident instances using a probability threshold to create instances for the subsequent classification and false positive reduction. As it is desired to include many candidates, so to increase the recall rate, we choose a lower probability threshold $T_p = 0.35$. This typically increases the false positive rate, but the subsequent module is devised to mitigate that.

### 3.4.4 Self-Supervised Instance Classification

The final step of the proposed pipeline is a self-supervised instance-based classification in order to detect nuclei that their representation falls out of the normal nuclei and their appearance is closer to background.

This ROI instance based classification is similar to FLAIR module but with two main differences: (1) it is performed in a global level and (2) there are no size-based criteria to select instances.

The hypothesis here is the following: so long as the majority of instances is correctly classified, the minority of instances that are false positives do not affect the ensemble learning, as they are statistically less significant. Moreover, if their representation is closer to the background, rather than the foreground then they are simply classified as false positives and are removed from the final segmentation output.

For feature extraction and classification, we use the *H* image and convert it to the HSI colorspace (as in NuSegHop). We apply feature extraction on each channel separately and concatenate them before the classifier's input. For features we opt for the first order statistics to learn the color characteristics, as well as the gray-level zone matrix which includes several features that capture the rough texture of the nuclei [33]. Here, we choose an SVM classifier with a radial basis kernel, to predict for each instance the probability of being a true positive.

Table 2: Summary of hyperparameters configuration in LG-NuSegHop, finetuned on a small subset of training images from MoNuSeg.

| Hyperparameter | Value |
|---|---|
| *Local Processing* | |
| $\lambda$ (Adapt. Thresh.) | 0.2 |
| $s$ (LAIR) | 0.7 |
| $\gamma$ (LAIR) | 0.1 |
| *NuSegHop* | |
| Energy $T_e$ | $10e-4$ |
| # Spectral Dimensions | 10 |
| XGB – # trees | 100 |
| XGB – Tree depth | 4 |
| XGB – Learning rate | 0.075 |
| *Global Processing* | |
| Blob Threshold (LoG) | 0.05 |
| $T_p$ | 0.35 |

## 4 Experimental Results

This section includes details of the experiments conducted in this work to validate the efficiency of our work, as well as potential areas of improvement. The datasets used for experiments and comparisons are briefly introduced in subsection 4.1, while the metrics for the quantitative analysis in 4.2. Additionally, our method is compared against other state-of-the-art methods, from unsupervised to weakly and fully supervised methods in 4.3. Furthermore, an ablation study is carried out in 4.4 to evaluate how different modules affect the performance, accompanied with visualization examples (4.4.1). In subsection 4.5 we discuss on the findings from the comparisons with the state-of-the-art.

### 4.1 Datasets

Three publicly available datasets are chosen to compare our proposed methodology.

**MoNuSeg** [18] The Multi-Organ Nuclei Segmentation (MoNuSeg) dataset includes 30 training image tiles of size 1000x1000 and magnification level 40x, comprising 21, 623 nuclei. Also, 14 testing imag tiles are available for benhmarking. The extracted tiles come from histological slides of breast, liver, kidney, prostate, bladder, colon, and stomach col-lected from The Cancer Genome Atlas(TCGA) [53]. Also, the samples collected come from different hospitals and patients. Therefore, it is a fairly diverse dataset across different aspects that challenges the generalization ability of the segmentation model. The Aperio ImageScope was used for digitizing the slides.

**CryoNuSeg** [39] This is the first H&E multi-organ dataset from frozen samples. Slides digitization from frozen samples involves a different process and therefore the nuclei appearance is different. This technique pertains intra-operative surgical sessions and its major benefit is that it can be performed rapidly. Yet, the requirement for a quick slide preparation, staining and digitization comes at the expense of the image quality. The dataset provides 30 digitized images of size $512 \times 512$, acquired at a $40\times$ magnification level. The slides come from 10 different organs (larynx gland, adrenal, lymph nodes, pancreas, skin, pleura, mediastinum, thyroid gland, thymus, testes) and there exist 7,596 annotated nuclei.

**CoNSeP** [9] This dataset includes 41 image tiles from 16 slides of patients with colorectal adenocarcinoma. 27 tiles are used for training and 14 for testing. The extracted tiles are of size $1000 \times 1000$. The Omnyx VL120 scanner was used for the slides digitization at a $40\times$ magnification level. Overall, 24,319 nuclei are annotated.

### 4.2 Evaluation Metrics

For performance evaluation, we use three different metrics, that have been commonly used in the literature. It is worth noting that nuclei segmentation is an instance-level segmentation problem. That is, a nucleus instance needs to be detected and then segmented. F1 score is the harmonic mean of the precision and recall. The F1 score regards nuclei segmentation as an instance detection problem, without taking into account the segmentation aspect. To complete our metrics, we also include the Aggregated Jaccard Index (AJI) metric [19] and the Panoptic Quality (PQ) [15]. These two metrics are more suitable for instance-level segmentation problems as they take into account both aspects. In particular, PQ calculates the detection quality (DQ), as well as the segmentation quality as a similarity measure with the ground truth. Dice similarity coefficient is also included in our comparisons to measure the segmentation performance.

### 4.3  Results & Comparisons

#### 4.3.1  Experimental Setup

To have a thorough understanding on the advantages and weaknesses of our work, we compare it against several state-of-the-art works with different levels of supervision. At first, we compare our method with other self-supervised methods on the MoNuSeg dataset, which use no labels from the target datasets. Another category is the weakly supervised methods that either use less training samples from the annotation masks or point annotations. Moreover, we include in our analysis a few popular fully supervised methods, so as to provide a thorough comparison of our work.

Before we delve into the comparisons, one aspect we would like to stress is that our method does not use any training data for parameters learning. Yet, since there are several hyperparameters (see Table 2) need adjustment, we use 6 out of the 30 training images randomly from the MoNuSeg dataset to finetune the LG-NuSegHop. This can be viewed as the validation set of our experiments. Hence, the hyperparameters are empirically finetuned using this validation set. At first, we adjust the local processing module, using the AJI and F1 score calculated from the intermediate (i.e. pseudolabel) output. Then, NuSegHop and global processing modules are adjusted together. For all the modules we tune the hyperparameters to maximize the AJI metric, whereas in the last module of the global processing set (self-supervised instance classification) we try to maximize the F1 score, since it is an ROI-wise classification task.

After the model is fixed, we test it out on the three testing datasets. This aims at testing how well our model generalizes to data with inherent discrepancies. However, we individually finetune LG-NuSegHop hyperparameters on each datasets using their a subset of their training data, in order to also test the performance when LG-NuSegHop is adapted to a certain domain.

#### 4.3.2  Performance benchmarking

At first glance, LG-NuSegHop has a competitive standing in comparison with other works, including also the fully supervised ones (Tables 3, 4 and 5). In the MoNuSeg dataset, it outperforms by large margins the self and weakly supervised works in terms of all the reported metrics, with a 0.651 AJI. It also has an impressive performance standing among the fully supervised works, including sophisticated models for nuclei segmentation, such as HoVer-Net [8] and CDNet [11] (see Table 3). However, other models such as NucleiSegNet [22] and CIA-Net [70] achieve higher performance in AJI, yet LG-NuSegHop achieves a competitive standing in detecting nuclei based on the F1 score (0.892 vs. 0.901). Comparing our method with GUSL model from the GL methodology, it achieves an AJI score of 0.673. The full supervision of GUSL helps to learn the more challenging distinctions between nuclei and background, which are hard to capture by a self-supervised method. Yet, in terms of F1 score, the detection performance is similar. Also, when fine-tuned using all the MoNuSeg training images, the performance increases slightly to the 0.658 AJI score and 0.892 F1 score. In terms of the Dice score our method also achieves the best per-formance among the weakly and self supervised methods by significant difference from the second leading performance (0.791 vs. 0.774).

Table 3: Performance benchmarking with self, weakly and fully supervised methods in the MoNuSeg dataset.

|  | AJI | F1 | Dice |
|---|---|---|---|
| *Self Supervised* | | | |
| DARCNN [13] | 0.446 | 0.5410 | - |
| Self-Attention [49] | 0.535 | - | 0.747 |
| CyC-PDAM [28] | 0.561 | 0.748 | - |
| Nucleus-Aware [51] | 0.593 | 0.759 | - |
| *Weakly Supervised* | | | |
| Partial Points [43] | 0.543 | 0.776 | 0.732 |
| Point Annotations [26] | 0.562 | 0.776 | 0.744 |
| BoNuS [27] | 0.607 | 0.780 | 0.767 |
| Cyclic Learning [69] | 0.636 | 0.774 | 0.774 |
| *Fully Supervised* | | | |
| U-Net [47] | 0.543 | 0.779 | - |
| RCSAU-Net [56] | 0.619 | 0.82 | - |
| HoVer-Net [8] | 0.618 | 0.826 | - |
| CDNet [11] | 0.633 | 0.831 | - |
| TopoSeg [12] | 0.643 | - | - |
| GUSL [63] | 0.673 | 0.886 | 0.803 |
| NucleiSegNet [22] | 0.688 | 0.813 | - |
| CIA-Net [70] | **0.691** | **0.901** | - |
| **LG-NuSegHop (Baseline)** | 0.651 | 0.887 | 0.778 |
| **LG-NuSegHop (Dom. Adapted)** | 0.658 | 0.892 | **0.791** |

Table 4: Performance benchmarking with weakly and fully supervised methods in the CryoNuSeg dataset.

|                              | AJI   | Dice  | PQ    |
|------------------------------|-------|-------|-------|
| *Weakly Supervised*          |       |       |       |
| BoNuS [27]                   | 0.431 | 0.693 | 0.399 |
| Partial Points [43]          | 0.410 | 0.682 | 0.357 |
| DAWN [67]                    | 0.508 | 0.804 | 0.476 |
| Pseudoedgenet [66]           | 0.321 | 0.620 | 0.306 |
| DoNuSeg [57]                 | 0.441 | 0.672 | 0.306 |
| *Fully Supervised*           |       |       |       |
| U-Net [47]                   | 0.469 | 0.697 | 0.403 |
| HoVer-Net [8]                | 0.526 | 0.804 | 0.495 |
| Swin-unet [3]                | 0.524 | **0.849** | 0.498 |
| CDNet [11]                   | 0.539 | 0.776 | **0.499** |
| **LG-NuSegHop (Baseline)**   | 0.545 | 0.703 | 0.419 |
| **LG-NuSegHop (Dom. Adapted)** | **0.567** | 0.723 | 0.479 |

In the CryoNuSeg, our method surpasses all the weakly supervised methods by large margins, achieving an AJI of 0.545 (Table 4). Comparing with the Dice coefficient and PQ, only the recently proposed DAWN [67] has a better performance. Besides, from the fully supervised category LG-NuSegHop achieves a higher AJI score from the state-of-the-art. Remarkably, even when it is not finetuned on the CryoNuSeg data (i.e. baseline model), LG-NuSegHop achieves a competitive performance in this dataset, where the acquisition process is considerably different than the standard H&E staining process. When finetuned on the same domain, the AJI performance increases further to 0.567. Overall, our method achieves a high AJI and PQ performance comparing to the other supervised methods. Full supervision in this dataset seems to boost more the segmentation performance, as one can see from the higher Dice score, comparing to the weakly or unsupervised methods.

In the third dataset under comparison, CoNSeP, all methods achieve a lower performance –compared to the other datasets–, since it has a large intra-class variance, where nuclei have quite different textures. Hence, this dataset is challenging for most methods. LG-NuSeg achieves an AJI of 0.422 without any finetuning (Table 5). Comparing to the weakly supervised methods it has a competitive performance across all metrics. Only the point annotations [52] and DAWN [67] methods perform better. Compared to the fully supervised works, the performance gap is larger. Yet, if we finetune the hyperparameters on the training images, the AJI score increases significantly to 0.461, surpassing most of the weakly supervised works and narrowing the gap with the fully supervised ones, such as U-Net. Considering all the metrics, in this dataset fully supervised methods achieve a higher performance in both segmentation and detection metrics. Therefore, one can infer that when the nuclei appearance variance is higher, supervision is crucial for both detecting and delineating the nuclei cells. This can be also observed in Fig. 6, where our method achieves a good segmentation result in MoNuSeg and CryoNuSeg, while in CoNSeP, for certain nuclei types it is more challenging to detect and segment them accurately without supervision or any domain adaptation.

Table 5: Performance benchmarking with weakly and fully supervised methods in the CoNSeP dataset.

|  | AJI | Dice | PQ |
| --- | --- | --- | --- |
| *Weakly Supervised* | | | |
| Pseudoedgenet [66] | 0.221 | 0.331 | 0.153 |
| BoNuS [27] | 0.354 | 0.651 | 0.380 |
| Partial Points [43] | 0.366 | 0.646 | 0.391 |
| Point Annotations [52] | 0.464 | 0.749 | 0.398 |
| DAWN [67] | 0.509 | 0.805 | 0.477 |
| *Fully Supervised* | | | |
| U-Net [47] | 0.499 | 0.761 | 0.434 |
| HoVer-Net [8] | 0.513 | **0.837** | 0.492 |
| CDNet [11] | **0.541** | 0.835 | **0.514** |
| Mulvernet [55] | 0.515 | 0.833 | 0.482 |
| **LG-NuSegHop (Baseline)** | 0.422 | 0.654 | 0.407 |
| **LG-NuSegHop (Dom. Adapted)** | 0.461 | 0.691 | 0.427 |

### 4.4 Ablation Study

As LG-NuSegHop is a pipeline consisting multiple processing steps and modules, we conduct an ablation study with different modules to demonstrate their efficacy and importance within the overall pipeline. Table 6 shows a few comparisons between the PQR and L,AB color conversion as a pre-processing step, as well as the contribution of the local processing modules. Table 7 at first compares a handcrafted feature extraction approach against NuSegHop. In turn, we progressively add the global processing modules to test the performance improvement.

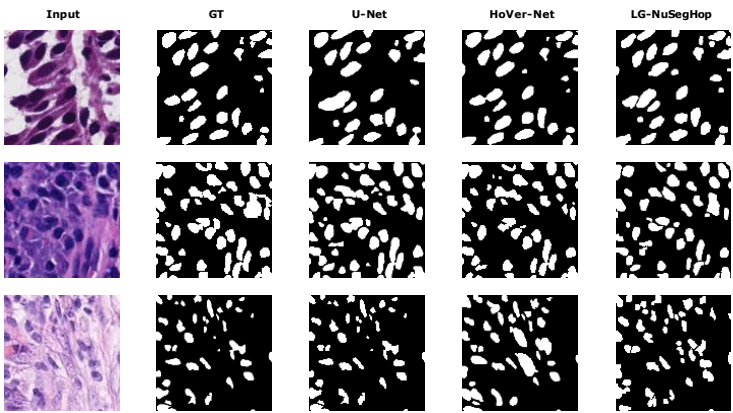

Figure 6: Visual comparisons of LG-NuSegHop with HoVer-Net and U-Net on patch examples from MoNuSeg (first row), CryoNuSeg (second row) and ConSeP (third row) datasets.

First of all, one can observe that our PQR pre-processing conversion helps both the detection and segmentation metrics. Also, the adaptive thresholding improves significantly the AJI over the non-adaptive one. Morphological post-processing is important to remove noisy instances and split nuclei and it is reflected from the large improvement of the F1 score (see Table 6). LAIR module also provides a small improvement in F1 score, by filtering some false positives in images, wherever it is more likely to have a high false positive rate. On the other hand, by replacing a handcrafted feature extraction in lieu of NuSegHop, all metrics drop significantly. Moreover, local maxima detection on NuSegHop's heatmap improves mainly the detection performance by recalling areas that indicate nuclei existence. It is also evident that both watershed and self-supervised instance classification help to improve mainly the detection aspect of the task, by reducing the false positive rate and delineate the nuclei more precisely (see Table 7).

Table 6: Ablation study on the MoNuSeg dataset with combinations of preprocessing and local processing modules.

| LAB | PQR | $T_o$ | $\hat{T}$ | Morph. Refin. | LAIR | AJI | F1 | DICE |
|---|---|---|---|---|---|---|---|---|
| ✓ | | ✓ | ✓ | | ✓ | 0.605 | 0.763 | 0.732 |
| | ✓ | | ✓ | ✓ | ✓ | **0.611** | **0.779** | **0.738** |
| | ✓ | ✓ | | | | 0.583 | 0.745 | 0.705 |
| | ✓ | | ✓ | | | 0.595 | 0.749 | 0.721 |
| | ✓ | | ✓ | ✓ | | 0.608 | 0.774 | 0.728 |

Table 7: Ablation study on the MoNuSeg dataset with different global processing modules. NuSegHop data-driven features are also compared with a set of handcrafted features for nuclei segmentation. All the pre-processing and local processing operations are kept to their best configuration.

| [68] | NuSegHop | LMD | Watershed | ROI Clasf. | AJI | F1 | DICE |
|---|---|---|---|---|---|---|---|
| ✓ | | | | | 0.622 | 0.813 | 0.740 |
| | ✓ | | | | 0.641 | 0.836 | 0.763 |
| | ✓ | ✓ | | | 0.647 | 0.875 | 0.769 |
| | ✓ | ✓ | ✓ | | 0.651 | 0.883 | 0.772 |
| | ✓ | ✓ | ✓ | ✓ | **0.658** | **0.892** | **0.778** |

#### 4.4.1 Qualitative Analysis

In the last part of our analysis, two visualization comparisons are provided. In Fig. 7, one can observe how the adaptive thresholding can help in segmenting local patches, where the bi-modal assumption is not very pronounced. The binarization performance is compared over a more conservative thresholding by choosing the intermediate point. It is demonstrated that the adjustment of the threshold about the intermediate point between the histogram peaks, helps to reduce the false positive areas upfront in LG-NuSegHop and thereby provides a less noisy pseudolabel to NuSegHop. Moreover, in Fig. 8, we illustrate a few examples across the three datasets and compare the instance self-supervised classification module operation in removing false positive instances as the last post-processing step of LG-NuSegHop. In MoNuSeg, our method achieves better results and in turn we can see that the false positive and negative rate is lower. CryoNuSeg yields a higher rate of false positives, which is mitigated from the global processing module. The ConSeP image has a higher false negative rate and hence the instance classification module does not have much effect.

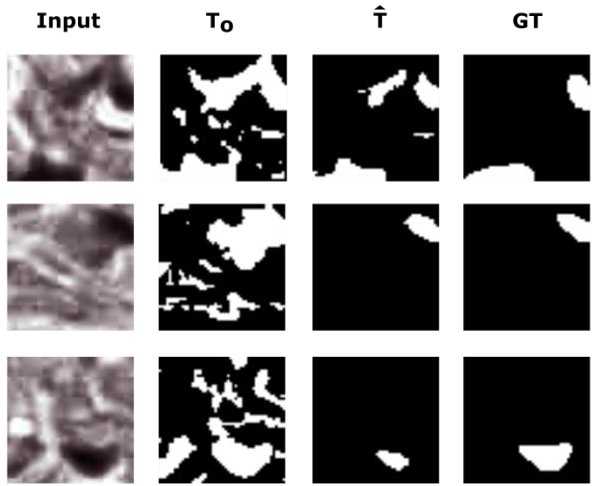

Figure 7: Illustrative examples of the adaptive filtering from the local processing module. $T_o$ is the intermediate point between the two peaks of the bi-modal distribution and the $\hat{T}$ the adapted threshold about $T_o$. The input patch is shown after the staining normalization.

## 4.5 Limitations & Discussion

From the quantitative and qualitative analysis, it is evident the importance and role of each individual module. In LG-NuSegHop pipeline other modules (e.g. instance classification) are meant to increase the detection accuracy and other the segmentation (i.e. NuSegHop). Overall, our proposed methodology achieves a very competitive performance among other self-supervised or weakly and fully supervised methods. It is worth noting that even without any finetuning on the target datasets, LG-NuSegHop outperforms most of the weakly supervised methods and can be compared to the state-of-the-art fully supervised ones. This seamless generalization ability provides a huge advantage of our method, since there is no need to form a training dataset or have a pathologists to annotate data. It can be deployed in a plug-n-play manner for nuclei segmentation and achieve a competitive performance in certain datasets (e.g. CryoNuSeg) or to provide pseudolabel to another self-supervised model.

As already mentioned, LG-NuSegHop is a pipeline that relies on hu-man's prior knowledge in solving the problem and requires certain assumptions to hold. From experimenting with the three diverse datasets, we can observe that our method shows a relatively lower performance on the CoNSeP dataset. This can be attributed to the fact that the local similarity assumption is very weak in this dataset. The nuclei instances have a large distance in shape and color, thus challenging for the local processing module to predict a less noisy pseudolabel for NuSegHop. In turn, it is also hard for the self-supervision to improve significantly over the pseudolabel prediction, resulting in a higher false negative rate.

From the experiments, fully supervised methods on the CoNSep dataset have a higher gap in performance both from the weakly supervised methods and the LG-NuSegHop one. Our method fits better into MoNuSeg and CryoNuSeg assumptions where the intra-class distance is relatively smaller and can yield a competitive performance even without any finetuning. One higher level conclusion that can be drawn from this comparison is that the nuclei segmentation problem on certain histology images can be solved solely by relying on clinical and biology prior knowledge using no or little supervision (Tables 3 and 4). Yet, when certain assumptions are not well met, full supervision is needed to learn the nuclei variability and achieve a higher performance (see Table 5). Another observation from the results and the comparison with different metrics is that our method is more effective on the detection aspect of the problem over the segmentation one, especially on the CryoNuSeg. The main reason is the appearance of the nuclei –faintly stained–, sourcing from the staining process. The lack of explicit full supervision makes harder for our method to accurately learn the boundaries distinctions for this type of staining. Thus, our method achieves a better Dice score, compared to the weakly supervised methods, but HoVer-Net and CDNet achieve better results due to their fully supervised training, which is crucial for this dataset in the accurate delineation of nuclei.

LG-NuSegHop can offer a high nuclei segmentation performance and generalize well with no specific domain adaptation. For example, although our method does not require any training from MoNuSeg, we use 6 images for hyperparameter finetuning. So, LG-NuSegHop is inherently adapted to this domain to a certain extend. Two recent works that have carried out domain adaptation from MoNuSeg (train) to CryoNuSeg (test), have achieved an AJI of 0.452 [23] and 0.484 [67]. For the same domain shift, LG-NuSegHop achieves an AJI of 0.545. Therefore, one can infer that supervision may not always be in favor of the cross domain generalization. We contend that for this task, biology priors and human insights play a pivotal role to mitigate the domain adaptation requirement.

From a complexity standpoint, LG-NuSegHop uses simple image processing operations pre and post the NuSegHop module with very low complexity. NuSegHop has a total number of parameters equal to 40*K* for feature extraction. Notably, the local processing pipeline and NuSegHop can be implemented in parallel to achieve a very short inference time. On the other hand, DL state-of-the-art solutions require several million of parameters and special equipment (e.g. large GPUs) for model deployment. For instance, as a reference point, HoVer-Net requires about 11.04 seconds to predict an image of size $1000 \times 1000$ using a GPU with 12GB memory [8]. On the same comparison, LG-NuSegHop requires about 9.38 seconds on average to predict a histology image of the same size, adopting a multi-thread implementation and deployed on a Intel Xeon CPU E5-2620 v3 at 2.40GHz.

As a final remark, it is also important to emphasize once more that within our pipeline every module is intuitive and transparent, including the feature extraction process in NuSegHop module. We underline this advantage as it is crucial for medical image solutions to be explainable to physicians, so that they can be effectively utilized in real clinical settings.

## 5  Conclusion

This work proposes the LG-NuSegHop pipeline for unsupervised nuclei segmentation from histology images. A novel feature extraction model named NuSegHop is introduced to learn the local texture. Regarding NuSegHop, several custom-made image processing modules are proposed to preprocess the input image, provide a pseudo label, and postprocess the predicted heatmap to increase the nuclei detection rate. Key advantages of our method are the generalization ability to unseen domains with inherent discrepancies and the small number of parameters. Every proposed module is intuitive and transparent, based on spe-cific biological priors of the problem. In future work, we will investigate ways to focus NuSegHop feature extraction on the nuclei boundaries, aiming to improve its segmentation performance. Moreover, another research direction is the incorporation of supervision into our current pipeline, in the challenging areas where the unsupervised approach has certain limitations, thus benefiting from the expert's knowledge.

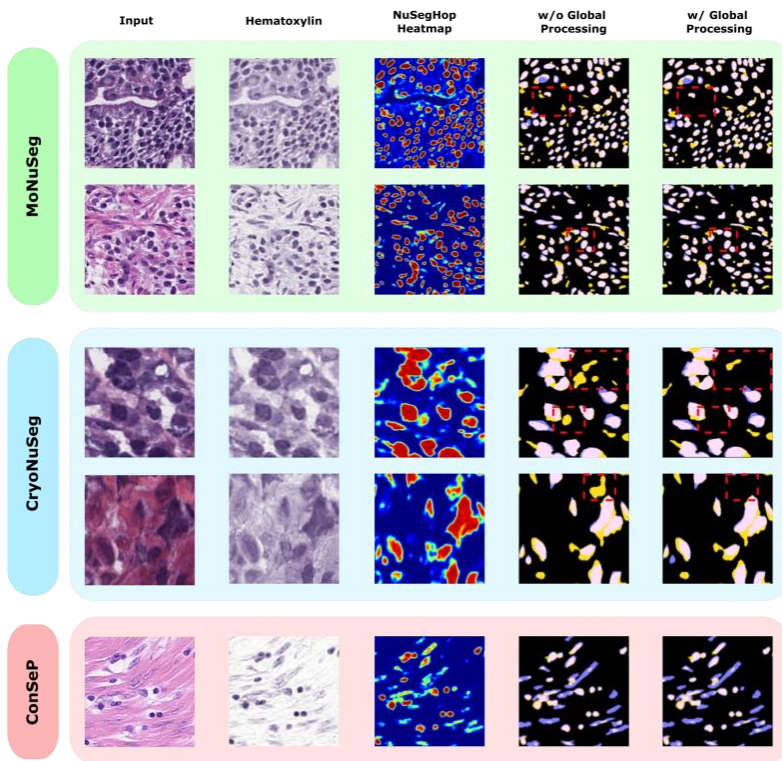

Figure 8: Visualization examples of the nuclei segmentation performance in three datasets. It is also compared the performance of the instance ROI classification within the global processing module. The true positive areas are marked with white, the false positives with yellow and the false negatives with blue. Red boxes highlight areas where the false positive removal is successful.